\documentclass[journal,comsoc,10pt]{IEEEtran}

\usepackage[T1]{fontenc}

\usepackage{cite}
\usepackage[colorlinks,
linkcolor=red,
anchorcolor=blue,
citecolor=green
]{hyperref}

\ifCLASSINFOpdf
\else
\fi
\usepackage{amsmath}
\usepackage{diagbox}

\usepackage{ amssymb }
\usepackage{amsfonts}
\usepackage{caption}
\usepackage{graphicx}
\usepackage{lettrine}
\usepackage{epstopdf}
\usepackage{textcomp,booktabs}
\usepackage[usenames,dvipsnames]{color}
\usepackage{colortbl}
\definecolor{mygray}{gray}{.9}
\definecolor{mypink}{rgb}{.99,.91,.95}
\definecolor{mycyan}{cmyk}{.3,0,0,0}
\usepackage{pifont}
\usepackage{subfig}
\usepackage{multirow}
\usepackage{url}
\usepackage{color}
\usepackage{threeparttable}
\usepackage{setspace}
\usepackage{lineno}
\usepackage{tabularx}
\usepackage[dvipsnames,usenames]{color}

\usepackage{bm}

\usepackage{algorithm} 
\usepackage{algorithmic} 

\usepackage{dblfloatfix}

\usepackage[dvipsnames,usenames]{color}
\usepackage[usenames,dvipsnames]{color}
\newcommand{\bfbl}[1]{\bf \color{blue}#1}

\definecolor{light-gray}{gray}{0.90}
\begin{document}
	\title{Semantic Communications using Foundation Models: Design Approaches and Open Issues }

	\author{Peiwen Jiang, Chao-Kai Wen, Xinping Yi, Xiao~Li, Shi Jin, and Jun Zhang
			\thanks{P. Jiang, X. Yi, X. Li and S. Jin are with the National
				Mobile Communications Research Laboratory, Southeast University, Nanjing
				210096, China (e-mail: PeiwenJiang@seu.edu.cn; 
xyi@seu.edu.cn; li\_xiao@seu.edu.cn; jinshi@seu.edu.cn).}
			\thanks{C.-K. Wen is with the Institute of Communications Engineering, National
				Sun Yat-sen University, Kaohsiung 80424, Taiwan (e-mail: chaokai.wen@mail.nsysu.edu.tw).}
			\thanks{J. Zhang is with the Department of Electronic and Computer Engineering,
Hong Kong University of Science and Technology, Hong Kong (e-mail: eejzhang@ust.hk).}}
	
	\maketitle
	\pagestyle{empty}  
	\thispagestyle{empty} 
%
%

\begin{abstract}
Foundation models (FMs), including large language models, have become increasingly popular due to their wide-ranging applicability and ability to understand human-like semantics. While previous research has explored the use of FMs in semantic communications to improve semantic extraction and reconstruction, the impact of these models on different system levels, considering computation and memory complexity, requires further analysis. This study focuses on integrating FMs at the effectiveness, semantic, and physical levels, using universal knowledge to profoundly transform system design. Additionally, it examines the use of compact models to balance performance and complexity, comparing three separate approaches that employ FMs. Ultimately, the study highlights unresolved issues in the field that need addressing.
\end{abstract}

	\section{Introduction}

\IEEEPARstart{S}{emantic} communication, vital for improving transmission efficiency in emerging mobile applications, relies heavily on state-of-the-art artificial intelligence (AI) models \cite{gunduz2022beyond}. Various AI techniques are employed in existing semantic methods to address different aspects. A domain adaptation method is proposed for multi-task systems to identify user requirements and assess task performance \cite{zhang2022unified}. However, the complexity of semantic metrics makes them unsuitable as derivable loss functions, necessitating reinforcement learning methods \cite{lu2021reinforcement}. A proper knowledge base (KB) for a specific content and task helps in extracting transmitted semantic features and restoring received ones. Most studies establish an implicit KB through joint source-channel coding \cite{bourtsoulatze2019deep,xie2020deep}. Some physical modules \cite{jiang2022deep,zhang2023scan,wu2022vision} are also redesigned to adaptively transmit semantic features according to changing physical channels and user requirements. Despite the demonstrated superiority of semantic communications, existing methods are primarily suited to specific content or tasks.

The field of Natural Language Processing (NLP) has experienced remarkable advancements due to innovations in AI technology. A pivotal development was the self-attention mechanism and Transformer-based neural network architectures. Pre-trained models, which learn universal language representations from extensive unsupervised data, became popular in the late 2010s, obviating the need for training new models from scratch. Recently, researchers discovered that enlarging models and training data notably enhances model capacity. Simultaneously, generative models have emerged as superior solvers for general tasks. Noteworthy examples of generative foundation models (FMs) include GPT-3.5 (ChatGPT) with 175 billion parameters and DALL-E with 12 billion parameters. The triumph of FMs presents opportunities for creating a universal semantic method suitable for any transmission scenario.

 	 \begin{figure*}[!t]
	\centering

	\includegraphics[width=6in]{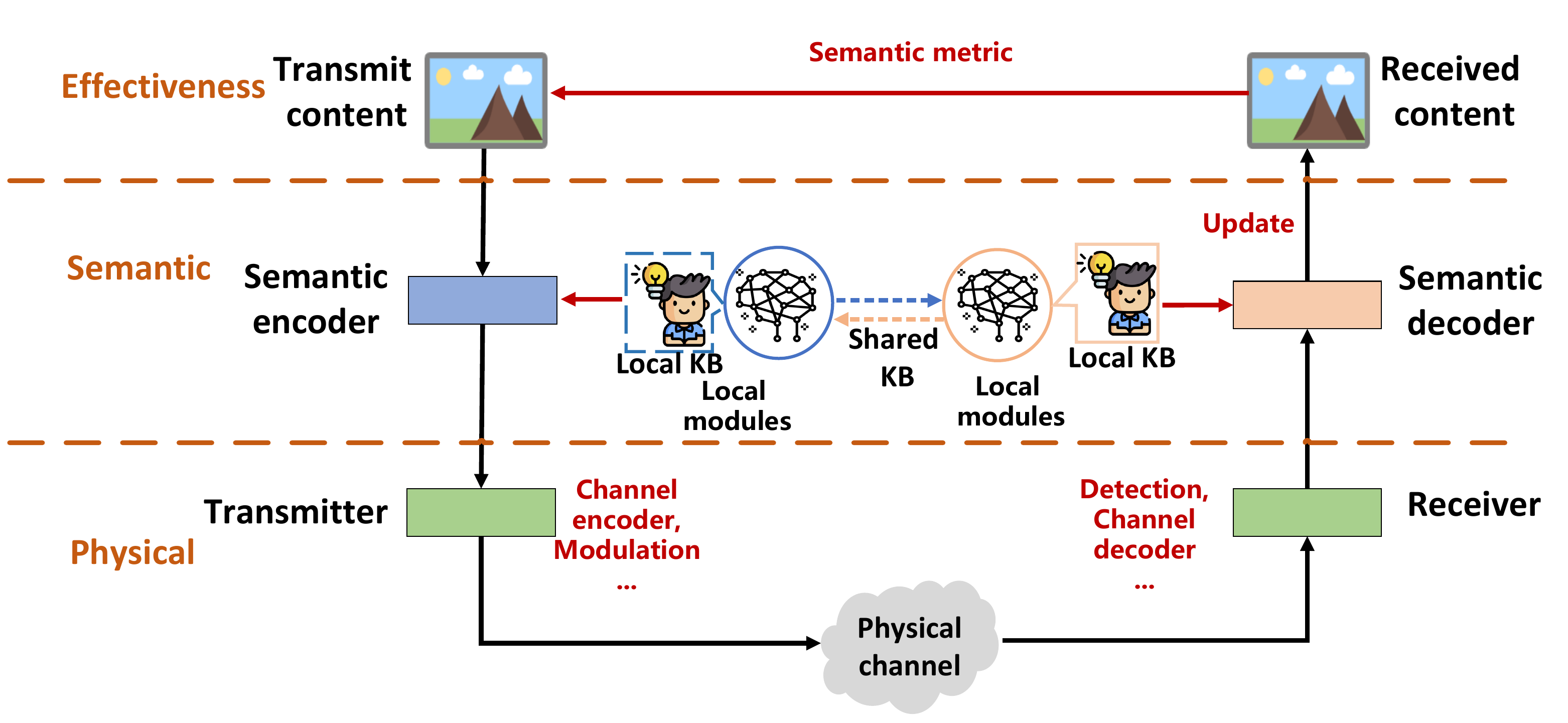}
	\caption{The conventional framework of semantic communication.}
	\label{Old}
\end{figure*}

Current research \cite{jiang2023large,shen2023large} guides the integration of FMs into semantic communications. However, there is a lack of discussion on leveraging FMs across multiple dimensions of semantic communications, especially considering the constraints of limited computational resources at the physical level. This article begins by outlining the current semantic framework, identifying its shortcomings, and then highlighting the potential advantages of integrating FMs. Subsequently, we provide an overview of FM-based semantic communications, encompassing innovative alterations at various levels. Although FMs revolutionize the semantic transmission framework, the sheer number of parameters still limits their applicability, and we elucidate three methods to emphasize the merits of compact model design. Building upon these efforts in FM-based semantic communications, we highlight several challenges that require resolution to establish a robust system.

The structure of this article is as follows: Section II elucidates the current semantic framework and the advantages of incorporating FM. Section III examines the ramifications of FMs on distinct levels of semantic transmission and underscores the necessity for compact model design. Section IV contrasts three approaches to utilizing FMs in semantic communications. Section V addresses unresolved issues in FM-based semantic communications. Finally, Section VI concludes the article.

\section{Semantic Communication meets FMs}
This section commences by introducing the current semantic communication framework and pinpointing its limitations. Subsequently, it delves into the transformations introduced by FMs in the realm of semantic communication.

\subsection{Current Semantic Communication Framework}
In contrast to conventional communication systems, semantic methods completely reshape the design, organizing it into three levels: effectiveness, semantic, and physical levels \cite{bao2011towards}, as illustrated in Fig. \ref{Old}.
\begin{itemize}
\item The {\bf effectiveness level} assesses transmission impacts and user requirements, necessitating task recognition and semantic metric calculations.

\item At the {\bf semantic level}, semantic coding techniques are utilized to extract transmitted semantic features and correct received ones using KBs. Local and shared KBs are essential, expressed either implicitly through trainable parameters or explicitly through transmitted examples.

\item The {\bf physical level} is responsible for transmitting semantic features and can be redesigned, as the transmission metric is no longer limited to the bit error rate. Innovative physical modules provide adaptive protection based on changing channel conditions and diverse tasks.
\end{itemize}

While existing semantic communication methods exhibit advantages in certain aspects, they also introduce new challenges. Semantic communications primarily focus on transmission effectiveness, making system design complex when dealing with multiple users and tasks. Most KBs are implicitly established through joint training, making them challenging to update. Additionally, fixed networks and trained semantic modules lack flexibility in adapting to varying physical channels. Consequently, more effective design methodologies are required.

 \begin{figure*}[!t]
	\centering

	\includegraphics[width=6in]{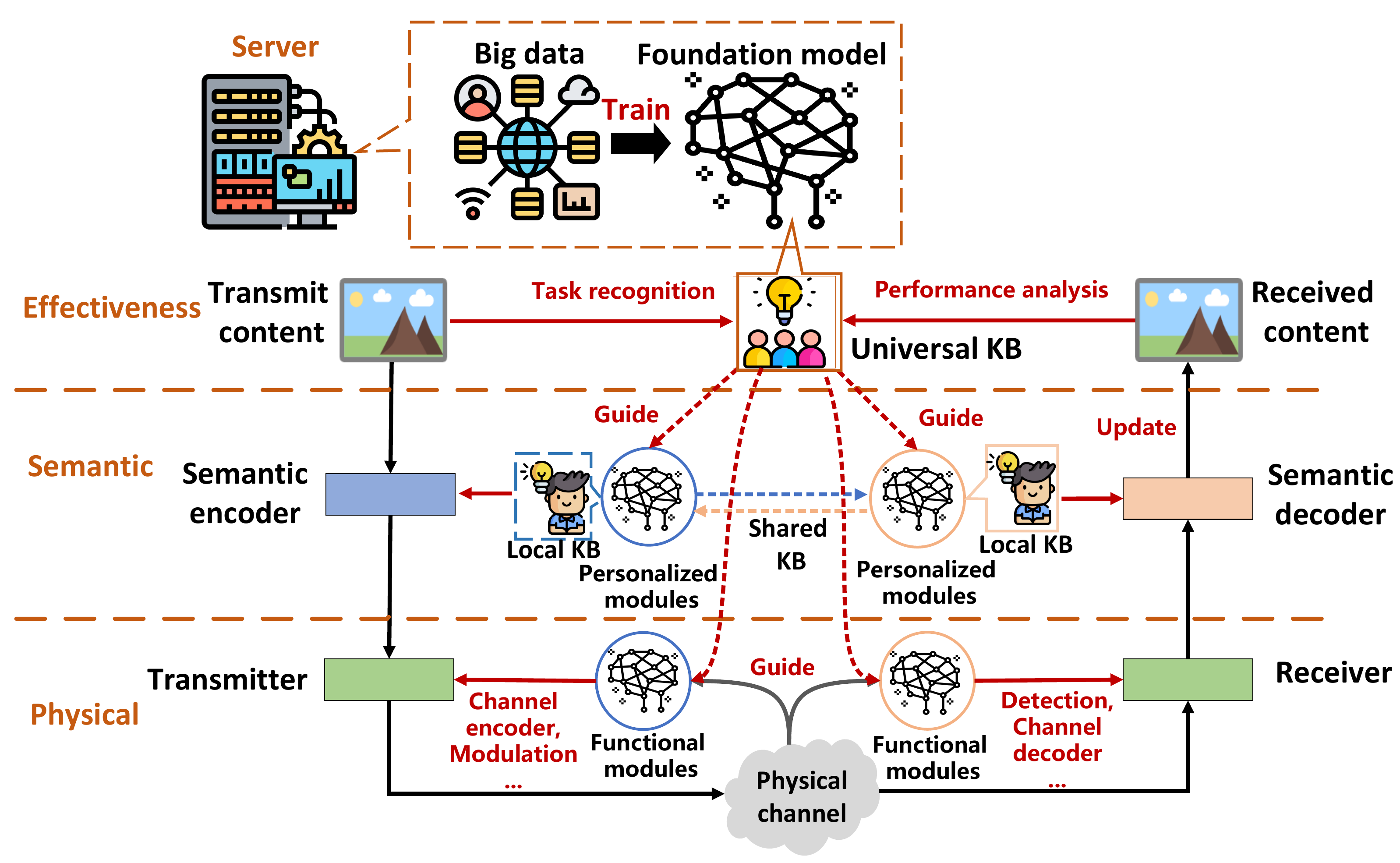}
	
	\caption{FM-enhanced semantic communication: collaborating modules across three levels with the universal KB. }
	\label{Conventional}
\end{figure*} 

\subsection{Integrating FMs into Semantic Communication}
FMs have demonstrated remarkable performance in comprehending and generating human-like text, surpassing smaller models by a significant margin. Consequently, these models have emerged as a highly promising trend in AI research. FMs revolutionize system design by enabling efficient literature analysis, generating novel hypotheses, and facilitating complex data interpretation. By amalgamating knowledge from various scenarios and tasks, FMs can serve as a versatile task solver, driving substantial advancements across various fields.

FMs are a natural fit for semantic communication since semantic methods rely on understanding transmitted content. Traditional semantic communication is only effective under specific contexts and tasks. For instance, training a semantic sentence transmission model on a textual dataset effectively handles syntax and semantics. However, these models fall short of replicating human capabilities because common sense and factual knowledge, encapsulated in billions of model parameters, must be learned from billions of training samples. Common sense and factual knowledge, often referred to as universal knowledge in FMs, are crucial for humans to navigate unfamiliar content and tasks. By harnessing FMs, semantic communications have the potential to become universally applicable to all content and tasks. 

However, certain concerns persist:
\begin{itemize}
    \item \textbf{High Computation and Memory Demands:} The computational and memory requirements of FMs pose obstacles to their integration into communication systems, particularly at the terminal's physical level.
    \item \textbf{Lack of Domain Specialization:} FMs may face challenges in expert fields \cite{ling2023beyond}, especially in future applications such as automatic driving, video conferencing, and extended reality, where domain-specific expertise is crucial.
    \item \textbf{Trustworthiness:} Trust is a critical issue in communication systems. FM-based semantic communication can generate flawless content from transmitted information, yet it can also produce different contents from the same transmitted information, making it challenging to detect fake content. Moreover, explaining FM behavior may prove to be an insurmountable task.
\end{itemize}

\section{Semantic Communication with FMs}
In addition to influencing various levels, FMs have inspired diverse design approaches aimed at enhancing semantic communication. Some methods directly utilize FMs to improve the performance of specific semantic modules, such as semantic segmentation, importance division, and reconstruction. Recognizing the roles played by FMs is crucial in designing a streamlined model to reduce complexity and boost performance. In this section, we will explore the contributions of FMs to existing semantic communication, encompassing:

Besides affecting different levels, FMs have inspired different design ways for enhancing semantic communication. Some approaches directly use a FM to enhance the performance of a semantic module, such as semantic segmentation, importance division, and reconstruction. Recognizing the roles played by FMs is important to design a compact model to reduce the complexity and enhance the performance. In this section, we will discuss what FMs provide to the existing semantic communication, including:

\begin{itemize}
	\item Universal KB with common sense and factual knowledge,

    \item Versatile analyzer for diverse user requirements at the effectiveness level,

	\item General interface for varied contents and applications at the semantic level,

	\item Innovative guidance for module design and optimization at the physical level.
\end{itemize}

\subsection{FM Enabling a Universal KB}
KBs are pivotal in semantic communications, typically falling into two categories: shared and local KBs. Local KBs at the transmitter and receiver are responsible for semantic extraction and reconstruction. The shared KB between the transmitter and receiver, for a specific transmitted content and user requirement, conserves transmission resources. In most existing semantic architectures, these KBs are implicitly embedded in the training parameters and are treated as black boxes. Consequently, novel adaptive semantic studies focus on locally updating parameters and subsequently sharing them with other terminals. Alternatively, an explicit approach to sharing the KB using content examples is also prevalent.

FMs introduce a novel perspective with a universal KB that incorporates common sense and factual knowledge. This implies that FMs can proficiently handle a broad spectrum of contents and tasks. Semantic communications grounded in FMs are viewed as a potential avenue to establish universal semantic systems. However, due to the high complexity and suboptimal expert performance of FMs, replacing local and shared KBs in current semantic communications with the universal KB directly is impractical. Since equipping all terminals with FMs is infeasible, local and shared KBs can be made \textbf{compact} to complement the universal KB. In Fig. \ref{Conventional}, at the semantic and physical levels, compact modules are depicted as personalized and functional modules. \textbf{Personalized modules} economize resources by removing task-unrelated knowledge from FMs while incorporating domain knowledge to enhance task performance. \textbf{Functional modules} economize resources by focusing on specific functions, such as semantized channel coding and modulation, while improving performance by safeguarding features according to the universal KB.

In summary, the universal KB from FMs presents an opportunity to establish a generalized semantic method for solving a wide array of tasks. Considering the practical system complexities and performance factors, techniques like knowledge distillation and domain specialization offer a means to seamlessly integrate universal, local, and shared KBs.

\subsection{FM at the Effectiveness Level}
At the effectiveness level, transmitted contents and user requirements are diverse and ever-changing. Therefore, FMs offer a universal solution for content recognition and transmission performance analysis.

 \begin{figure*}[!h]
	\centering

		\includegraphics[width=7in]{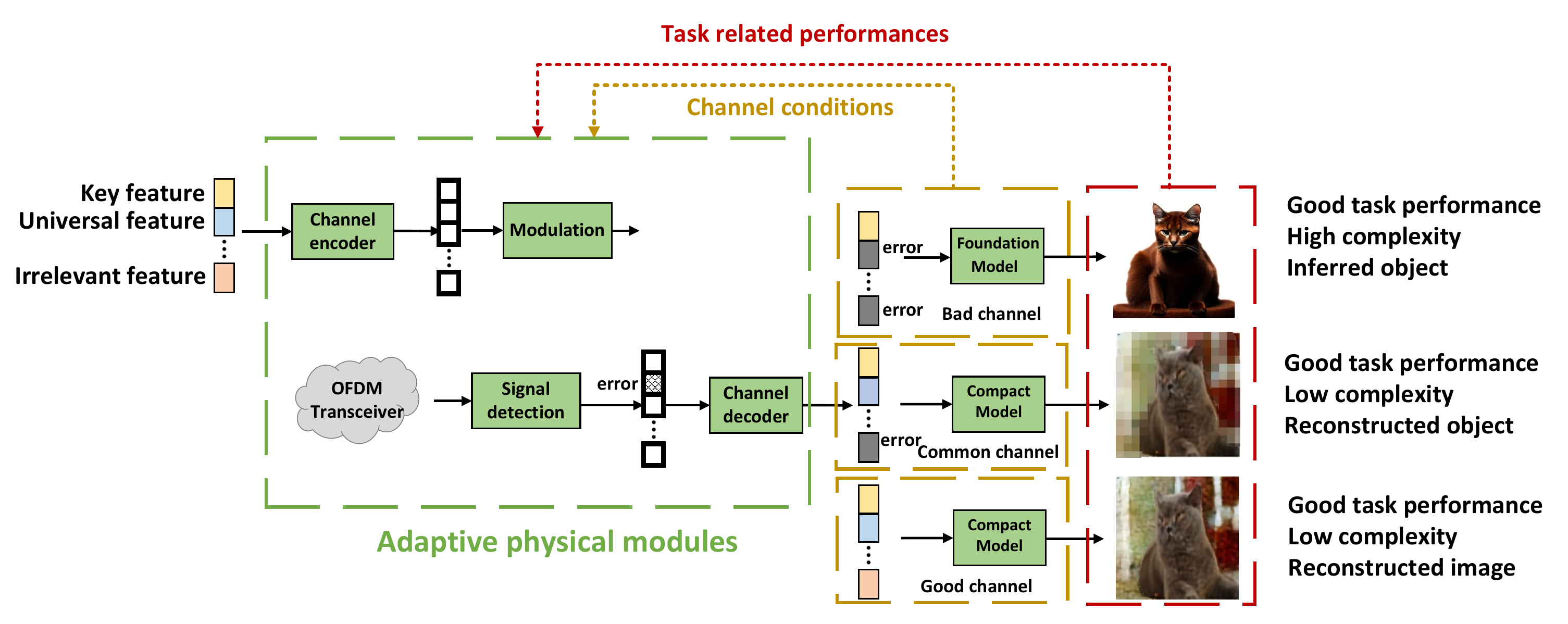}

	\caption{Adapting the physical module to user requirements and channel conditions: inferring transmission object with a complex FM under poor channel conditions and transmitting object honestly with a compact model under good channel conditions.}
	\label{Phy}
\end{figure*}

\textbf{Task Recognition:} With the advancement of AI-based communication technologies and applications, numerous AI models with various roles have emerged. For instance, in \cite{shen2023large}, GPT is leveraged to orchestrate multiple AI models within a communication system. Additionally, FMs can demonstrate their utility in adapting to different communication scenarios, consistently recommending suitable AI models or sets of models as transmission contents evolve.

\textbf{Performance Analysis:} FMs offer a universal approach to assess semantic performance. Current semantic metrics also rely on large models like BERT to measure semantic similarity. However, these metrics are tailored to specific content categories or tasks. A communication system equipped with numerous metrics can become complex and challenging to design. Novel large generative models can handle various semantic modalities, including text and images, and generate scores based on input content and requirements. With a substantial universal KB, FMs can score received content akin to human evaluation and guide communication system optimization.

In summary, FMs possess the capability to recognize and address problems comprehensively rather than focusing on a specific mission. This versatility benefits the effectiveness level, which requires a holistic analysis of transmission impacts.

\subsection{FM at the Semantic Level}
The semantic level plays a pivotal role in the semantic transmission framework by extracting and restoring semantic features based on KBs. Unlike conventional methods, semantic approaches can omit irrelevant content or reconstruct damaged portions using related semantic features. However, existing methods at the semantic level are typically tailored for specific targets and may struggle with diverse transmission data. FMs offer a universal interface for various semantic applications, and we discuss two well-studied examples: semantic segmentation and semantic reconstruction.

\textbf{Semantic Segmentation:} Semantic segmentation is widely applied in semantic communications because it allows explicit division of source content into different semantic parts. From a task performance perspective, only specific semantic segments need to be transmitted, directly conserving transmission bandwidth. In \cite{wang2023seggpt}, semantic segmentation can segment content using text control, making it adaptable to changing tasks and capable of outputting the required part from the source content.

\textbf{Semantic Reconstruction:} Semantic reconstruction leverages semantic correlations to repair missing or damaged semantic elements. Training a generator is a common approach in existing methods. However, small generators may excel in specific tasks but struggle with inconsistency and quality issues because they rely heavily on a powerful KB to compensate for the damaged parts. Large generative models, such as diffusion models, can generate high-quality content from any input information and are popular in tasks like text-to-image generation. In \cite{grassucci2023generative}, a diffusion model is used to restore transmitted semantic features, even with only outlines of required objects. This adaptability allows image transmission to adjust to changing channel conditions, and the large generative model can effectively reconstruct various types of transmitting information, including vital segments, contours, and descriptive text.

In summary, the semantic level enhanced by FMs possesses a universal capability for handling any transmitted content and task. These methods alleviate the burden of continuous model updates, resulting in more user-friendly outcomes. Furthermore, once the transmission task remains consistent over an extended period, FMs can be simplified and personalized to meet specific requirements.

\subsection{FM at the Physical Level}

 \begin{figure*}[!h]
	\centering

	\includegraphics[width=5.5in]{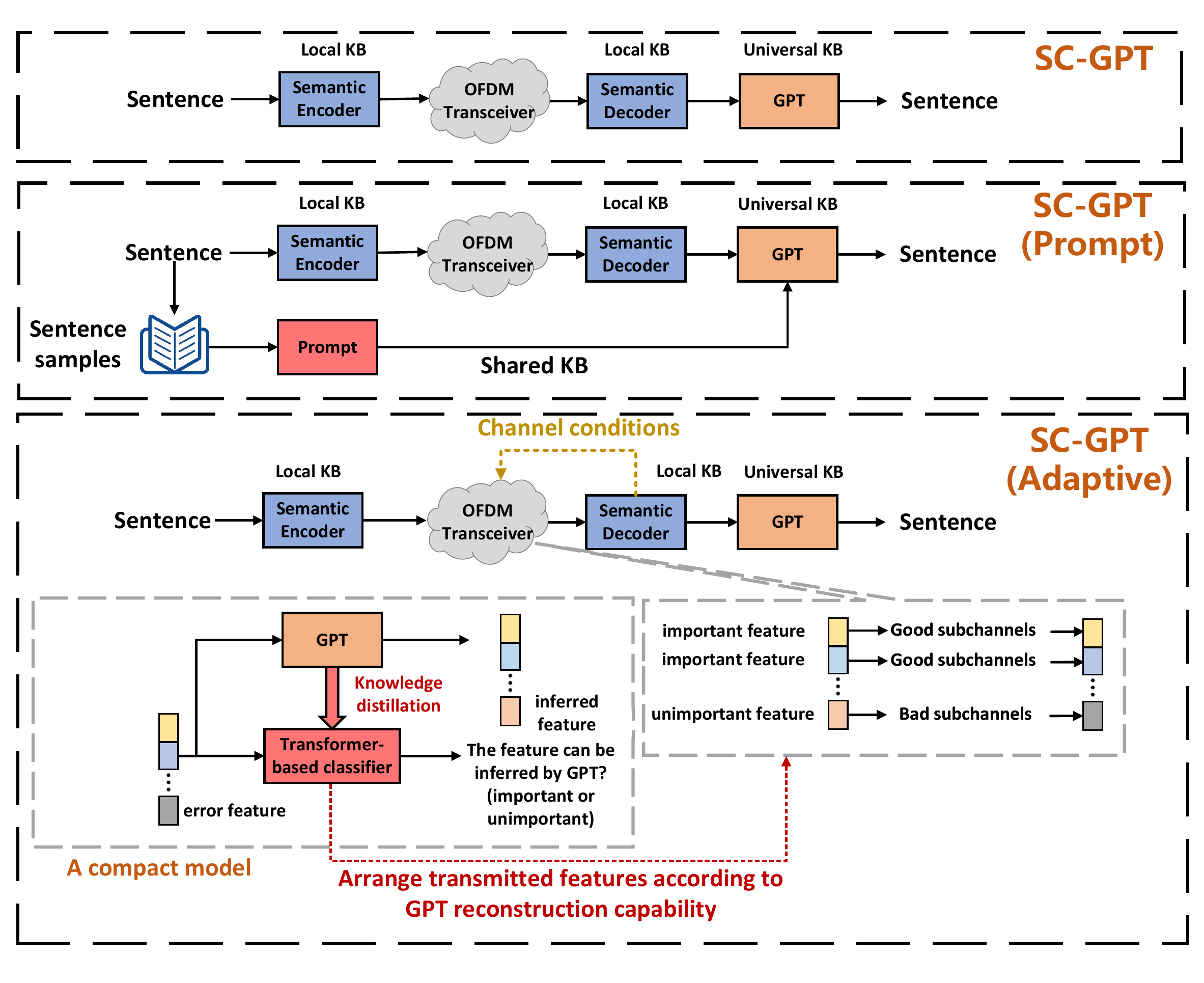}

	\caption{Comparison of three different methods utilizing FMs. }
	\label{Me}
\end{figure*}

After segmentation and extraction at the semantic level, the transmitted features undergo distortion in physical channels and are safeguarded by physical modules \cite{10183791}. In the context of image transmission for object recognition, different transmitted features play distinct roles under varying channel conditions:

 \begin{itemize}
     \item \textbf{Key Features:} These are the most critical features directly impacting task performance and cannot be restored from other features. Key features may include object descriptions and contours in an image or subjects and predicates in a sentence. In the example shown in Fig. \ref{Phy}, the key feature might be a sitting cat.

     \item \textbf{Universal Features:} These can only be predicted by an FM thanks to a large universal KB. In an image, universal features can be inferred as widely accepted components to fill in missing parts, as illustrated in Fig. \ref{Phy} under poor channel conditions. In a sentence, universal features might correspond to commonly known collocations for incorrect words.

     \item \textbf{Task-Relevant and Irrelevant Features:} These features are tailored to a specific task based on local KBs. Task-relevant features can be repaired using a simple model, while task-irrelevant features are disregarded. As shown in Fig. \ref{Phy}, a basic AI model can restore the transmitted object from key and universal features for an object recognition task. Under good channel conditions, all features are well transmitted, and the entire image is reconstructed.
 \end{itemize}

 With FMs, transmitted features can be highly compressed into only key features under poor channel conditions, which is a common approach in existing methods. However, transmitting only key features without considering changing physical channels may not be ideal. FMs can reconstruct the image from key features, but the received object may not be an accurate reconstruction of the source content, and the hardware complexity is significantly increased. When channel conditions are not extremely poor, a compact model similar to existing semantic methods might be a better choice. Such models provide an honest representation of the object, require lower complexity, and involve a slightly higher number of transmitted features. In fact, under good channel conditions, all features can be transmitted, and FMs are not always necessary to conserve resources.

 To accommodate varying channel conditions and requirements, physical modules should be adaptive and aware of semantics. These functional modules collaborate with FMs and other semantic methods to transmit the required semantic features, ensuring the desired quality is achieved.

\section{Three Approaches Utilizing FMs}

In this section, we introduce three semantic communication frameworks built upon FMs. Initially, we examine various methods of incorporating existing semantic coding (SC) and GPT-based text reconstruction. Subsequently, we present simulation results to validate the efficacy of these introduced modules and compare the performance of the different approaches.

 \begin{figure*}[!h]
	\centering

	\subfloat[ ]{
		\includegraphics[width=3.5in]{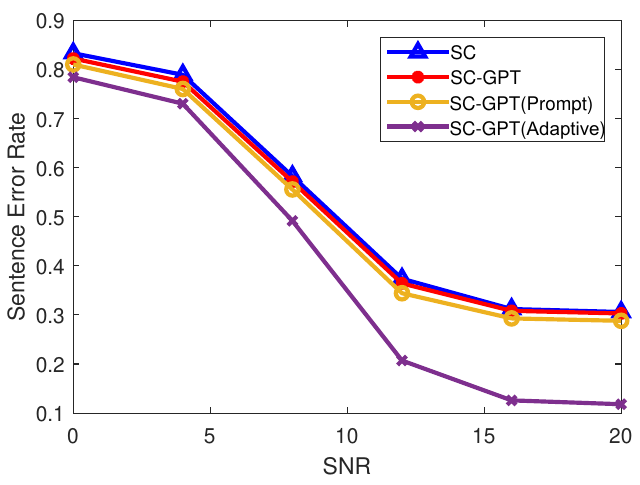}}
	\subfloat[ ]{
		\includegraphics[width=3.5in]{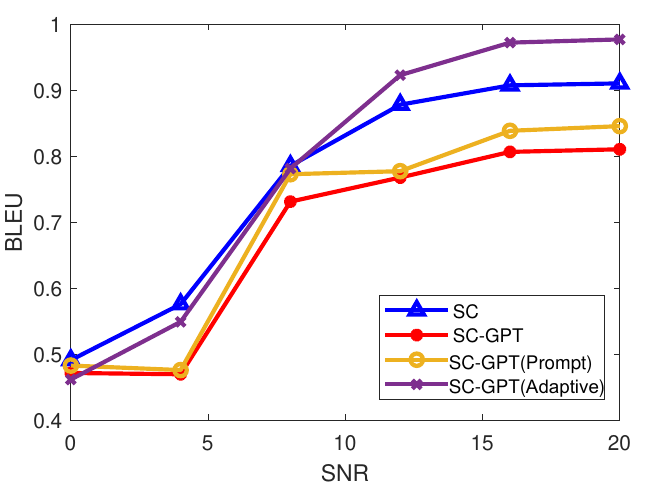}}
	
	\caption{Performances of different SC-GPT methods: (a) Sentence error rate. (b) BLEU.}
	\label{Results}
\end{figure*}

\subsection{Frameworks and Settings}

\subsubsection{General Framework}
We examine an uplink communication system where a user terminal sends a sentence to a base station (BS). Uplink transmission faces challenges due to limited antennas, sending power, and computational resources at the user terminal, making it a bottleneck for applications like video streaming. To alleviate this pressure, FMs are exclusively deployed at the BS to enhance reconstruction performance.

The fundamental transmission modules are set up and trained as follows:

\textbf{Semantic Coding:} Sentence transmission relies on a Transformer-based joint source-channel coding, following the architecture in \cite{jiang2022deep} for testing.

\textbf{GPT-based Text Reconstruction:} GPT-3.5 (ChatGPT) is utilized, commonly known for its versatility in various NLP tasks. We prompt this model with `{\tt Fix the errors in this sentence:...}'.

\textbf{OFDM Transceiver:} All methods are integrated into an OFDM system using 16-QAM modulation. Each block consists of eight OFDM symbols, each symbol containing 64 subcarriers. The first symbol serves as a pilot, while the following symbols carry data. The channel is modeled with three paths and exhibits an exponential power profile, with varying conditions between OFDM blocks.

\textbf{Dataset and Training Process:} Sentences are sourced from the European Parliament, ranging from 4 to 30 words in length. The training dataset includes 100,000 sentences, while the testing dataset comprises 10,000 sentences. The semantic encoder-decoder is trained with a learning rate of 0.0001.

A sentence at the user terminal is encoded by an SC encoder and transmitted using an OFDM transceiver. The SC decoder at the BS then outputs the received sentence. This received sentence is checked and corrected by GPT if it contains errors. This process implies that only erroneous sentences require GPT's assistance, and the overall complexity is similar to SC when the channel quality is good.

\subsubsection{Three Different Approaches}
Fig. \ref{Me} illustrates three approaches for integrating semantic coding and GPT-based text reconstruction.

\textbf{SC-GPT:} In this approach, GPT is employed to directly repair the output of the semantic decoder. This method can be easily implemented without additional training.

\textbf{SC-GPT(Prompt):} Recognizing that transmitted content typically concentrates on a specific domain, a simple model is used to summarize the content from sentence examples. This summary is sent in advance to prompt GPT. However, this approach increases overhead. Furthermore, as GPT lacks knowledge about transmission errors, both incorrect and correct sentence pairs are used to prompt GPT for in-context learning.

\textbf{SC-GPT(Adaptive):} This approach introduces a classifier at the transmitter to identify feature importance based on GPT's reconstruction capability. It is supported by importance-aware resource allocation to enhance GPT's reconstruction performance. The classifier is Transformer-based and uses the same encoder as semantic coding, along with a dense layer using the sigmoid activation function. The output dimension matches the number of semantic features. For each sentence, a transmitted feature is randomly disrupted, and GPT attempts to reconstruct the received sentence. If the disrupted feature can be repaired by GPT, it is labeled as \textbf{unimportant}, with the classifier's output value set to 0; otherwise, it is labeled as \textbf{important}, with the output value set to 1. The training process of the physical model can be viewed as a knowledge distillation process from GPT. Subsequently, the physical model assigns important features to good subchannels and the others to bad subchannels.

\subsection{Simulation Results}
	\begin{table*}[!h]
	\centering	
	\footnotesize
	\caption{The sentence examples received by different SC methods.  }
	\begin{tabular}{|>{\sf}c|l|p{12.5cm}|}    %
		\hline

\multicolumn{3}{|c|}{SC and SC-GPT}\\ \hline
\multirow{3}{*}{1}& SC& let me also remind you that over 90 of that budget is spent on projects that {\bfbl focus} member states \\ \cline{2-3}

&SC-GPT& let me also remind you that over 90 of that budget is spent on projects that {\bfbl focus on} member states \\ \cline{2-3}

&SC-GPT(Prompt)& let me also remind you that over 90 of that budget is spent on projects that benefit member states \\ \hline

\multirow{3}{*}{2}

&SC& first and {\bfbl crucial} there is a message for the council of ministers\\ \cline{2-3}

&SC-GPT& first and foremost there is a message {\bfbl in the commission that must be relayed to the ministers}\\ \cline{2-3}
&SC-GPT(Prompt)& first and foremost there is a message for the council of ministers\\ \hline
\multicolumn{3}{|c|}{SC-GPT(Adaptive)}\\ \hline
\multirow{6}{*}{1} & Proposed classifier (bold)& they have {\bfbl made} and are {\bfbl making considerable efforts}  in the {\bfbl fight} against {\bfbl corruption and organized crime}\\ \cline{2-3}
&SC without classifier& they {\bfbl very} made and are making considerable {\bfbl even} in the fight against {\bfbl there} and organised crime\\ \cline{2-3}

&SC-GPT& they have made and are making considerable
{\bfbl even} in the fight against terrorism and organized crime\\  \cline{2-3}
&SC with classifier& they have made and are making considerable {\bfbl even to} the fight {\bfbl the} corruption and organised crime\\ \cline{2-3}

&SC-GPT(Adaptive)& they have made and are making considerable efforts  in the fight against corruption and organized crime\\

		\hline
	\end{tabular}
	\label{T2}
\end{table*}

As depicted in Fig. \ref{Results}(a), we compare the sentence error rates of different methods. The SC-GPT methods exhibit an ability to correct more erroneous sentences than traditional SC. However, repairing wrong sentences can be challenging, particularly if their meanings have been altered. When compared to SC, SC-GPT only corrects a limited number of sentences. SC-GPT(Prompt) outperforms SC-GPT thanks to the summary and transmission examples, which aid in error correction. Ultimately, SC-GPT(Adaptive) offers the best performance by protecting features that GPT cannot repair.

Fig. \ref{Results}(b) presents BLEU performance for the competing methods. SC-GPT exhibits worse BLEU performance than SC when the SNR is low. This is because altered meanings in incorrect sentences can mislead GPT, leading to even poorer BLEU performance for such sentences. However, as the SNR increases, SC-GPT corrects more sentences and outperforms SC in terms of BLEU score. Similar patterns are observed in SC-GPT(Prompt). SC-GPT(Adaptive) significantly outperforms SC-GPT(Prompt) but still lags slightly behind SC when the SNR is low, particularly when key features are disrupted.

Table I offers examples showcasing the performance of the three methods. Errors are highlighted in blue. A comparison of SC and SC-GPT demonstrates GPT might introduce extra words based on context to correct missing words like ``focus on.'' SC-GPT(Prompt) emphasizes error correction rather than sentence rewriting, addressing this issue. Then, the effect of physical module is shown to explain the performance of SC-GPT(Adaptive). The proposed classifier identifies words challenging for GPT to correct. In this instance, SC with the classifier has a comparable number of error words as SC without it, while SC-GPT(Adaptive) effectively corrects errors.

In summary, the physical module demonstrates its potential to balance complexity and performance. For example, SC requires 4.9 million parameters, while the proposed physical module requires 2.5 million parameters. This simplicity becomes evident when compared to GPT, which has 175 billion parameters. Directly removing unimportant words at the transmitter to save bandwidth would necessitate GPT's involvement in every transmission, resulting in unacceptable complexity. Allocating unimportant words to poor subchannels means that the large-sized GPT can be omitted when channel conditions are favorable.

\section{Open Issues}
While FMs have achieved significant success and initial attempts have demonstrated their effectiveness in semantic communications, challenges remain, particularly in terms of computation and memory consumption. Big models, especially at the physical level, face limitations in practical application. Additionally, although big models provide a universal KB capable of addressing changing requirements, they lack domain knowledge, leading to inefficiencies in specific application scenarios. Further research should focus on intelligently harnessing the capabilities of big models.

\subsection{Establishing, Updating, and Cooperating between Local, Shared, and Universal KBs}
Existing semantic communication methods rely on local and shared KBs, embedded in model weights through joint training, to conserve transmission bandwidth. Adapting these trained models to changing content and tasks necessitates updates. While FMs offer the potential for revolutionary semantic communication design through a universal KB, the challenges of establishing and updating this universal KB are substantial. Retraining FMs can take months on powerful servers, rendering it impractical for edge servers and terminals. To address this, more feasible approaches include training adapters with fewer weights or using prompters. However, the complexities of running and fine-tuning large models persist, making them impractical for all terminals. Consequently, local and shared KBs carried by compact models should complement the universal KB, emphasizing domain specificity and adaptation rather than serving as compressed versions.

\subsection{Compact and Personalized Models to Strengthen Big AI-based Semantic Communications}
Existing semantic communications are task-specific at the effectiveness and semantic levels. Incorporating big models can create a universal semantic communication framework for all tasks. However, universal systems may lack domain knowledge for diverse users with evolving needs. Allocating a dedicated big model to each user is infeasible due to limited resources. Instead, FMs can serve as a general interface alongside personalized models at the effectiveness and semantic levels. These personalized models distill task-related KBs from big models, enhancing their suitability for specific domains. Establishing multiple compact and personalized models enables seamless adaptation to varying user requirements without frequent adjustments to big models.

\subsection{Compact and Functional Models to Accelerate Big AI-based Semantic Communications}
At the physical level, conventional modules primarily play functional roles such as channel coding, modulation, and resource allocation. While FMs can provide valuable reference information for transmission, including importance order, they cannot be directly applied to physical modules due to computational and time constraints. Compact models can be developed to mimic FMs but focus solely on specific physical functions, such as subchannel arrangement. These simplified models can be significantly smaller than the original large models while still cooperating with them through joint training. Enhancing all physical modules with simple and functional models can inherit the merits of big models and make the system feasible for various terminals.

\subsection{Compact and Explainable Models to Establish Trustworthy Semantic Communications}
Critical issues related to robustness, explainability, and proper evaluation can undermine the trustworthiness of existing semantic communications. FMs offer solutions for enhancing robustness, but their numerous parameters and powerful generative capabilities pose challenges for explanation and evaluation. Using explainable models to provide post-hoc explanations for FM behavior is a viable approach. These models analyze inputs, intermediate results, and outputs to shed light on the behavior of FMs. Additionally, performance evaluation may depend on evaluation models that consider various measures, including subjective scores and task performances.

\section{Conclusions}
This article has explored the advantages of integrating FMs into semantic communications. The challenges of limited practical resources, expert-level performance, and trustworthiness pose significant obstacles to system design. We have examined the impacts of FMs at various system levels and compared three FM-based methods. These approaches incorporate small-size local models to enhance task performance and alleviate computational demands. In conclusion, we have addressed key concerns across different system levels when harnessing the potential of FMs.

	\bibliographystyle{IEEEtran}
	\bibliography{bibtex0320}

\end{document}